\documentclass{IEEEcsmag}

\usepackage[colorlinks,urlcolor=blue,linkcolor=blue,citecolor=blue]{hyperref}
\expandafter\def\expandafter\UrlBreaks\expandafter{\UrlBreaks\do\/\do\*\do\-\do\~\do\'\do\"\do\-}
\usepackage{upmath,color}

\usepackage{xcolor}
\usepackage{xspace}
\usepackage{subcaption}
\usepackage{pifont}
\usepackage{graphicx}
\usepackage{graphbox}
\usepackage{subcaption}
\usepackage{dblfloatfix} 
\usepackage{epstopdf}
\usepackage[htt]{hyphenat}
\let\rfoot\relax
\usepackage{fancyhdr}
\usepackage{setspace}
\usepackage{relsize}
\usepackage{textcomp}
\usepackage{circledsteps}
\usepackage{url}
\usepackage[htt]{hyphenat}
\usepackage[super]{nth}
\usepackage{comment}
\usepackage{transparent}
\usepackage{nicematrix,booktabs}
\usepackage{balance}

\usepackage{calc}
\usepackage[T1]{fontenc}
\usepackage{amsmath}
\usepackage[cmintegrals]{newtxmath}
\usepackage[utf8]{inputenc}
\usepackage{tabularx}
\usepackage{array}
\usepackage{multirow}
\usepackage{booktabs}
\usepackage[para]{threeparttable}
\usepackage{colortbl}
\usepackage{multirow}
\usepackage[framemethod=tikz]{mdframed}
\newcommand{\red}[1]{\bgroup\color{red}#1\egroup\xspace}
\newcommand{\blue}[1]{\bgroup\color{blue}#1\egroup\xspace}
\newcommand{\teal}[1]{\bgroup\color{teal}#1\egroup\xspace}
\newcommand{\green}[1]{\bgroup\color{green}#1\egroup\xspace}
\tikzset{
  vertical align/.style={
    baseline=-.5*(height("$+$")-depth("$+$"))
  }
}
\newcommand*{\circleNum}[1]{
  \tikz[baseline=(char.base)]{
    \node[shape=circle,fill,inner sep=0.9pt] (char) {%
        \textcolor{white}{\scriptsize{#1}%
      }%
    };%
  }%
}

\usepackage[numbers,super,sort&compress]{natbib}

\usepackage[T1]{fontenc}
\usepackage{lipsum}
\usepackage[normalem]{ulem}
\makeatletter
\def\SOUL@hlpreamble{%
\setul{\dimexpr\dp\strutbox-2pt}{\dimexpr\ht\strutbox+\dp\strutbox-2pt\relax}
\let\SOUL@stcolor\SOUL@hlcolor
\SOUL@stpreamble
}
\makeatother

\newcommand\khlc[1][yellow]{
  \bgroup
  \markoverwith{\textcolor{#1}{\rule[-.5ex]{1pt}{2.5ex}}}
  \ULon
}

\newcommand{\todo}[1]{\bgroup\color{white}\textbf{\khlc[black]{TODO: [#1]}}\egroup\xspace}
\newcommand{\fixme}[1]{\bgroup\color{red}\textbf{\khlc{FIXME: [#1]}}\egroup\xspace}
\newcommand{\pointer}[1]{\bgroup\color{white}\textbf{\khlc[red]{POINTER: [#1 is working here]}}\egroup\xspace}

\newcommand{\reviewer}[1]{\bgroup\color{blue}#1\egroup\xspace}
\newcommand{\ranswer}[1]{\bgroup\color{red}#1\egroup\xspace}

\let\labelindent\relax
\usepackage{enumitem}
\usepackage[framemethod=tikz]{mdframed}
\usepackage[numbers,sort&compress]{natbib}
\tikzset{
    vertical align/.style={
        baseline=-.5*(height("$+$")-depth("$+$"))
    }
}

\usepackage{calc}
\usepackage{amsmath,algorithm,algpseudocode}
\usepackage{cases}
\algrenewcommand\algorithmicindent{0.5em}
\makeatletter
\algrenewcommand\ALG@beginalgorithmic{\footnotesize}
\makeatother
\makeatletter
\renewcommand{\Function}[2]{%
  \csname ALG@cmd@\ALG@L @Function\endcsname{#1}{#2}%
  \def\jayden@currentfunction{#1}%
}
\newcommand{\funclabel}[1]{%
  \@bsphack
  \protected@write\@auxout{}{%
    \string\newlabel{#1}{{\jayden@currentfunction}{\thepage}}%
  }%
  \@esphack
}
\makeatother

\usepackage{pythonhighlight}
\definecolor{codegray}{rgb}{0.5,0.5,0.5}
\lstdefinestyle{pythonStyle}{
  basicstyle=\tiny\ttfamily\footnotesize\linespread{0.5},
  basewidth = {.54em},
  commentstyle=\color{codegray},
  frame=single,
  language=Python,
  stepnumber=1,
  numbers=left,
  numbersep=5pt,
  numberstyle=\tiny\color{codegray},
  tabsize=1,
  showspaces=false,
  showstringspaces=false,
  breaklines=false,
  mathescape,
  keywordstyle={\color{black}},
  emph={Load, GraphPre, BatchPre, Aggr, Trans},
  emphstyle={\bfseries\color{orange}},
  moredelim=**[is][\color{red}]{~}{~},
  moredelim=**[is][\color{blue}]{<}{>},
  moredelim=**[is][\color{orange}]{@}{@},
  literate={\\~}{{\textasciitilde}}1
  {\\<}{{\unichar{"003C}}}1
  {\\>}{{\unichar{"003E}}}1
  {\\@}{{\unichar{"0040}}}1
}

\usepackage{tabularx}
\usepackage{array}
\usepackage{multirow}
\usepackage{booktabs}
\usepackage[para]{threeparttable}
\usepackage{colortbl}
\usepackage{multirow}

\makeatletter
\def\hlinewd#1{%
\noalign{\ifnum0=`}\fi\hrule \@height #1 %
\futurelet\reserved@a\@xhline}
\makeatother
\usepackage{hhline}
\newcolumntype{C}{>{\centering\arraybackslash}X}

\usepackage[framemethod=tikz]{mdframed}
\usepackage{marginnote}
\usepackage{pdfcomment}
\mdfsetup{
  hidealllines=true,
  innerleftmargin=2pt,
  innerrightmargin=2pt,
  innertopmargin=2pt,
  innerbottommargin=2pt,
  leftmargin=-2pt,
  rightmargin=-2pt,
  skipabove=-2pt,
  skipbelow=-2pt,
  backgroundcolor=blue!20}

\newlength{\markerHeight}
\newlength{\markerMargin}
\newlength{\linespace}
\newlength{\linedepth}

\sbox0{yf}
\definecolor{mylime}{RGB}{205, 220, 57}
\definecolor{mygreen}{RGB}{60, 200, 0}
\definecolor{myblue}{RGB}{0, 51, 204}

\colorlet{soulred}{red!20}
\colorlet{soulgreen}{green!20}
\colorlet{soulblue}{blue!20}

\usepackage{enumitem}
\setlistdepth{5}
\renewlist{itemize}{itemize}{5}
\setlist[itemize,1]{label=$\bullet$}
\setlist[itemize,2]{label=$\circ$}
\setlist[itemize,3]{label=$\ast$}
\setlist[itemize,4]{label=-}
\setlist[itemize,5]{label=$\cdot$}

\makeatletter
\def\SOUL@hlpreamble{%
\setul{\dimexpr\dp\strutbox-2pt}{\dimexpr\ht\strutbox+\dp\strutbox-2pt\relax}
\let\SOUL@stcolor\SOUL@hlcolor
\SOUL@stpreamble
}
\makeatother

\jtitle{This manuscript is an extended version of the original paper accepted by IEEE Micro.}
\pubyear{2025}

\begin{document}

\sptitle{THEME ARTICLE: CACHE COHERENT INTERCONNECTS AND RESOURCE DISAGGREGATION TECHNIQUES}

\title{CXL-GPU: Pushing GPU Memory Boundaries with the Integration of CXL Technologies}

\author{Donghyun Gouk$^{*}$, Seungkwan Kang$^{\dagger}$, Seungjun Lee$^{\dagger}$, Jiseon Kim$^{*}$, Kyungkuk Nam$^{*}$, Eojin Ryu$^{*}$, Sangwon Lee$^{*}$, Dongpyung Kim$^{*}$, Junhyeok Jang$^{*}$, Hanyeoreum Bae$^{*}$, and Myoungsoo Jung$^{*\dagger\ddagger}$
}
\affil
{
\\
\\$^{*}$Next-Generation Silicon and Research Division, \textbf{Panmnesia, Inc.}, Daejeon, South Korea
\\$^\ddagger$Advanced Product Engineering Division, \textbf{Panmnesia, Inc.}, Seoul, South Korea
\\$^\dagger$KAIST, Daejeon, South Korea
}

\begin{abstract}
\looseness=-1
This work introduces a GPU storage expansion solution utilizing CXL, featuring a novel GPU system design with multiple CXL root ports for integrating diverse storage media (DRAMs and/or SSDs). We developed and siliconized a custom CXL controller integrated at the hardware RTL level, achieving two-digit nanosecond roundtrip latency, the first in the field. This study also includes speculative read and deterministic store mechanisms to efficiently manage read and write operations to hide the endpoint's backend media latency variation. Performance evaluations reveal our approach significantly outperforms existing methods, marking a substantial advancement in GPU storage technology.
\vspace{-8pt}

\end{abstract}

\maketitle

\chapteri{L}arge-scale deep learning models, such as large language models and mixtures of experts, have become prevalent across various computing domains. However, their memory demands often far exceed the capacity of accelerators like GPUs. For instance, while models with 1 billion parameters require approximately 16$\sim$24 GB of GPU memory for loading and training, models exceeding 100 billion parameters are increasingly commonplace.

To overcome these limitations, researchers have explored various approaches that leverage storage solutions or external memory systems. One notable example is GPU Direct Storage (\textit{GPUDirect} \cite{nvidia2022gpudirectstorage}), which enables direct mapping of the GPU's PCIe base address register (BAR) to a target SSD by altering the storage stack. This approach facilitates the management of large-scale model parameters by utilizing the substantial storage capacity of SSDs. However, GPUDirect's adoption remains limited due to its complexity. Treating the SSD as a block device introduces challenges, including the need for detailed management of the file system and the mismatch in I/O granularity between memory and block operations. Furthermore, GPUDirect requires manual intervention for storage and memory operations (e.g., \texttt{cuFileWrite}), complicating the copy-then-execute programming model and deterring widespread usage.

In contrast, \emph{unified virtual memory} (UVM \cite{nvidia2022cuda}) provides a more straightforward solution by enabling shared virtual memory access for both CPU and GPU. This approach integrates GPU and host memory into a unified address space, allowing seamless data access. UVM automatically handles memory allocation and migration based on demand, simplifying the management of large datasets. As a result, UVM has become a widely adopted technique for deep learning frameworks like TensorFlow and DGL.

\begin{figure}
  \centering
  \includegraphics[width=\linewidth]{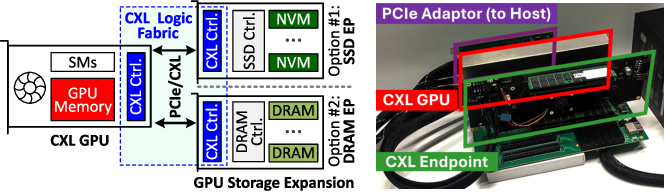}
  \begin{subfigure}{\linewidth}
    \begin{tabularx}{\textwidth}{
      p{\dimexpr.5\linewidth-2\tabcolsep-1.3333\arrayrulewidth}
      p{\dimexpr.5\linewidth-2\tabcolsep-1.3333\arrayrulewidth}
    }
      \vspace{-5pt} \caption{Overview.} \label{figs:overview-and-framework1} &
      \vspace{-5pt} \caption{AIC-level prototype.} \label{figs:overview-and-framework2}
    \end{tabularx}
  \end{subfigure}
  \vspace{0pt}
  \caption{High-level view of GPU storage expansion.}
  \label{figs:overview-and-framework}
  \vspace{0pt}
\end{figure}

In this study, we present the \emph{compute express link} (CXL) as a transformative technology for GPU storage expansion. Figure \ref{figs:overview-and-framework1} illustrates the high-level framework for the proposed GPU storage expansion approach. CXL leverages the PCIe physical data link to map devices, referred to as \emph{endpoints} (EPs), to a cacheable memory space accessible by the host. This architecture allows compute units to directly access EPs using standard memory requests. Unlike JEDEC's DDR-standard DRAM interfaces, which rely on synchronous communication, CXL enables asynchronous communication with compute units, akin to block storage. This flexibility supports EP implementation using diverse storage media such as non-volatile memory (NVM)-based SSDs \cite{jung2022hello} and DRAM \cite{kim2023smt,maruf2023tpp}.

Despite its potential, integrating CXL for GPU storage expansion presents a significant challenge due to the absence of a native CXL logic fabric and subsystem in GPUs to support DRAM/SSD EPs as memory expansion devices. To address this limitation, we have developed a CXL hardware layer stack and implemented it through a custom CXL controller designed at the hardware \textit{register-transfer level} (RTL), integrated within the CXL logic fabric. Building on this, we propose a GPU architecture featuring multiple CXL root ports equipped with these controllers, each capable of interfacing with DRAM/SSD EPs via a host bridge. This work represents the first demonstration of a silicon-based CXL controller achieving sub-two-digit nanosecond round-trip latency, signifying a major advancement in high-speed memory expansion technology.

We further enhanced the design of our CXL controller to improve GPU storage expansion performance. While achieving the fastest CXL round-trip latency to date, the storage media in EPs remains slower than local GPU memory. To mitigate this latency gap, we propose two key strategies for GPU-EP interaction: \emph{speculative read} (SR) and \emph{deterministic store} (DS).

The SR mechanism leverages a module in our controller to manage load instructions, utilizing features from CXL 2.0. It anticipates target addresses by pre-sharing them with DRAM/SSD EPs, enabling these EPs to prefetch target pages before the actual requests arrive. This approach significantly reduces the latency impact of the storage media. To avoid overloading EPs with SR requests, we integrate the CXL quality of service (QoS) telemetry feature to monitor EP status and regulate SR traffic, ensuring a balance between performance and system load. For scenarios involving slower EP write performance, such as with NVM-based SSDs, the DS mechanism addresses the variability in write speeds. It performs concurrent writes to both GPU memory and the SSD EP, immediately completing store requests from compute resources. Since GPU workloads are predominantly read-oriented, this ``fire-and-forget'' strategy proves effective for most applications. However, in cases with intensive write activity or high SSD write latencies (referred to as tail cases), the DS mechanism identifies the SSD EP as a bottleneck. In such instances, our controller temporarily buffers incoming data in GPU memory, deferring their transfer to the SSD EP. This ensures that the GPU can sustain the fire-and-forget approach for the majority of store operations, delivering improved performance across diverse workloads.

To evaluate the performance of our CXL hardware framework, we fabricated the CXL controller using state-of-the-art silicon technology. Integration with the GPU was achieved using Vortex \cite{tine2021vortex}, a RISC-V-based general-purpose GPU, where our controller's RTL design was embedded. The effectiveness of our GPU storage expansion approach was thoroughly tested with custom-designed add-in-card (AIC) devices, as shown in Figure \ref{figs:overview-and-framework2}. Comprehensive testing was conducted using various storage media and GPU storage expansion techniques through RTL-level hardware simulations. The results demonstrate that our GPU storage expansion approach significantly outperforms the UVM strategy and a commercial EP prototype controller \cite{kim2023smt}, achieving 2.36$\times$ and 1.36$\times$ higher performance, respectively.

The key contributions of this work are as follows:

\begin{itemize}
  \item[{\ieeeguilsinglright}] \emph{Design of CXL-integrated GPU}: We propose a GPU architecture featuring a CXL root port and internal modifications, enabling direct access to memory expanders without host intervention.
  \item[{\ieeeguilsinglright}] \emph{Demonstration of a silicon-based CXL controller}: We present the development of a low-latency CXL silicon stack and its integration into GPU hardware at the RTL level, enabling high-speed memory expansion.
  \item[{\ieeeguilsinglright}] \emph{Speculative read and deterministic store mechanisms}: The SR mechanism anticipates target addresses to allow endpoints to prefetch target pages, while the DS mechanism manages concurrent writes to GPU memory and SSD EPs, optimizing performance.
\end{itemize}


\section{MEMORY MANAGEMENT IN GPU AND CXL TECHNOLOGIES}
\subsection{Data Movement Management in GPUs}

\noindent \textbf{Copy-then-execute model.}
GPU memory is crucial for storing not only model parameters but also additional data such as training metadata, gradients, and intermediate computation buffers. The memory requirements for large-scale deep learning models are often about eight times larger than the memory needed to store their parameters alone, frequently exceeding the capacity of modern GPUs, which are typically limited to a few tens or hundreds of GB. To address this limitation, various strategies have been proposed, including parallelism techniques that distribute workloads across multiple GPUs. While these approaches alleviate the memory burden on individual GPUs, they do not fully resolve the challenge of accommodating all necessary data within GPU memory.

\begin{figure}
  \centering
  \begin{subfigure}[]{.31\linewidth}
    \vspace{0pt}
    \includegraphics[width=\linewidth]{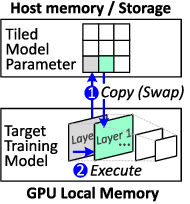}
  \end{subfigure}
  \begin{subfigure}[]{.66\linewidth}
    \vspace{0pt}
    \includegraphics[width=\linewidth]{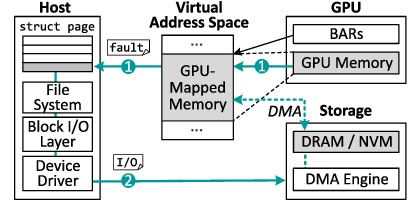}
  \end{subfigure}
  \begin{subfigure}{1\linewidth}
    \renewcommand*{\arraystretch}{0.3}
    \begin{tabularx}{\linewidth}{
      p{\dimexpr.32\linewidth-1.33\tabcolsep}
      p{\dimexpr.68\linewidth-1.33\tabcolsep}
      }
      \vspace{5pt} \caption{Data move.}\label{figs:copy-then-execute} &
      \vspace{5pt} \caption{Direct DMA for GPU and Storage.}\label{figs:direct-dma}
    \end{tabularx}
  \end{subfigure}
  \vspace{5pt}
  \caption{Data movement management in GPUs.}
  \vspace{3pt}
\end{figure}

Consequently, GPUs often rely on host memory and storage, necessitating frequent data transfers between the CPU and GPU. Figure \ref{figs:copy-then-execute} illustrates the ``copy-then-execute'' programming model commonly used for large-scale training. In this model, parameters are divided into smaller segments, or tiles, corresponding to different layers of the training model. This segmentation enables the swapping of layer output vectors (\blue{\circleNum{1}}), allowing subsequent layers to process with their assigned parameters (\blue{\circleNum{2}}). However, the frequent data transfers required in this model introduce significant performance overheads and complicate GPU usage. Users must explicitly manage data structures, handle tiled data transfers, and coordinate data migration, making the process both cumbersome and inefficient, ultimately posing a barrier to broader GPU adoption.

\noindent \textbf{Unified virtual memory.}
NVIDIA and AMD provide runtime software support for \textit{unified virtual memory} (UVM) in their GPU architectures \cite{nvidia2022cuda,amd2022rocm}. UVM enables both the CPU and GPU to access data via a shared pointer within a unified virtual address space. When the GPU encounters a memory access request for data not present in its memory, a cache miss and page fault occur. This triggers a PCIe interrupt, notifying the host. The host-side runtime software then allocates a new page in GPU memory, transfers the required data to this page, and updates the GPU. Memory within the virtual address space is thus allocated on demand. In addition, UVM automatically manages the GPU memory budget by replacing older pages with newly migrated data, ensuring efficient utilization. These features have made UVM a widely adopted solution in frameworks such as TensorFlow, DGL, and OneAPI.

However, while UVM enhances GPU accessibility, it can introduce performance bottlenecks. The primary cause is the substantial latency associated with host runtime intervention to resolve GPU page faults.

\noindent \textbf{Direct DMA for GPU and storage.}
UVM faces limitations in memory capacity, as it relies on host-side DRAM, which is itself constrained. Alternative approaches aim to expand GPU memory by utilizing block storage, significantly increasing overall memory capacity. Techniques such as \textit{GPU Direct Storage} (GPUDirect \cite{nvidia2022gpudirectstorage}) and NVMMU \cite{zhang2015nvmmu} share a common principle: granting the storage DMA engine access to the GPU's memory space to enable direct data transfers between GPU and storage.

As shown in Figure \ref{figs:direct-dma}, GPUDirect and NVMMU achieve this by mapping the GPU's BAR to the virtual address space and creating a \texttt{struct page} for GPU memory within the host kernel. This configuration allows the host's file system to share the GPU-mapped memory with storage, enabling storage devices to perform DMA operations directly to GPU memory. However, these approaches are limited by their reliance solely on the storage DMA engine for data transfers. Consequently, when the GPU encounters an on-demand page fault (\teal{\circleNum{1}}), the host runtime must translate all fault requests into storage I/O requests (\teal{\circleNum{2}}), resulting in overheads comparable to those seen in UVM.

\subsection{CXL Memory Expander}

\noindent \textbf{CXL memory interface.}
The CXL protocol is an open-standard interface designed to incorporate memory resources from \textit{endpoint} (EP) devices into the system's cacheable memory space \cite{cxl2023cxl}. It is structured around three sub-protocols: \textit{CXL.cache}, \textit{CXL.io}, and \textit{CXL.mem}. The CXL.cache protocol is responsible for maintaining cache coherence across various computing resources, ensuring data consistency when shared memory is accessed by multiple processors. This mechanism is critical to prevent mismatches or stale data in systems relying on shared memory spaces. The CXL.mem protocol facilitates reading and writing operations over the PCIe transport, using packetized communication through CXL flits. By enabling asynchronous memory operations, this protocol overcomes the rigid timing constraints inherent in traditional JEDEC-compliant memory interfaces, offering greater flexibility for memory access. The CXL.io protocol is similar to PCIe in its functionality, supporting device enumeration and managing bulk input/output (I/O) communication tasks. One notable use case of the CXL protocol is the CXL memory expander. This device has been highlighted for its ability to address memory limitations in computing systems, offering a pathway to improved scalability and mitigating memory-related bottlenecks \cite{maruf2023tpp}.

\begin{figure*}
  \begin{minipage}{.74\linewidth}
    \centering
    \vspace{0pt}
    \begin{subfigure}{.76\linewidth}
      \vspace{0pt}
      \includegraphics[width=\linewidth]{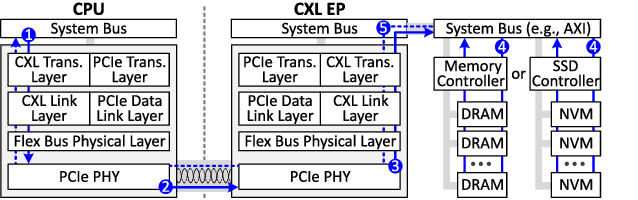}
    \end{subfigure}
    \hspace{-15pt}
    \begin{subfigure}{.19\linewidth}
      \vspace{0pt}
      \includegraphics[width=\linewidth]{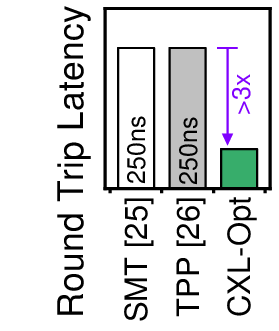}
    \end{subfigure}
    \begin{subfigure}{1\linewidth}
      \renewcommand*{\arraystretch}{0.3}
      \begin{tabularx}{\linewidth}{
        p{\dimexpr.79\linewidth-1.33\tabcolsep}
        p{\dimexpr.18\linewidth-1.33\tabcolsep}
        }
        \vspace{-5pt} \caption{Latency analysis.}\label{figs:latency-analysis} &
        \vspace{-5pt} \caption{Latency.}\label{figs:ctrl-perf}
      \end{tabularx}
    \end{subfigure}
    \vspace{-3pt}
    \caption{End-to-end latency of CXL.}
  \end{minipage}
  \hspace{8pt}
  \begin{minipage}{.235\linewidth}
    \vspace{0pt}
    \includegraphics[width=\linewidth]{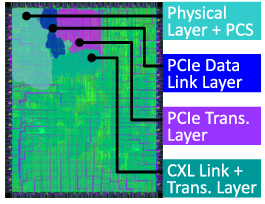}
    \vspace{5pt}
    \caption{Silicon layout.} \label{figs:floorplan-silicon}
  \end{minipage}
\end{figure*}

\noindent \textbf{Detailed end-to-end latency analysis.}
Many industry vendors have developed prototypes or proofs-of-concept for CXL memory expanders that leverage CXL 2.0. However, as of the present, commercially available products or in-depth latency information from an end-user perspective remains scarce. This subsection aims to detail the components contributing to CXL's end-to-end latency.

CXL's latency is distinctive from the JEDEC/DDR standards, which are characterized by precise timing specifications. Instead, CXL latency encompasses the full round-trip path across all relevant layers. Figure \ref{figs:latency-analysis} provides an in-depth look at the hardware layer stack, extending from the host to an EP. A memory request from the host CPU is converted into a CXL flit at the transaction layer, which supports the functionalities and interfaces required by each sub-protocol. This flit progresses to the link layer, which oversees the flow of communications, including buffering and acknowledgments. The flit is then passed to the Flex Bus physical layer (\blue{\circleNum{1}}), which sends it to the target EP based on factors such as link speed and lane configuration (\blue{\circleNum{2}}). For accurate management of CXL subprotocols, the EP must integrate these layers as well. After navigating the EP's layer stack, the flit returns to the system bus (\blue{\circleNum{3}}), reaching either a memory controller or an SSD controller. The data is then retrieved from the storage medium (\blue{\circleNum{4}}) and sent back to the host CPU's system bus, traversing the hardware layers once again (\blue{\circleNum{5}}). The PCIe transaction and link layers are crucial components due to their roles in device enumeration, configuration, and data management.

The structure and function of this hardware layer stack play a critical role in the determination of round-trip latency. Although market products are yet to be released, studies conducted by organizations like Samsung \cite{kim2023smt} and Meta \cite{maruf2023tpp} have examined their prototypes, reporting a latency of 250ns.

\noindent\textbf{CXL with an SSD integration.}
The asynchronous operation capabilities of CXL.mem enable support for a variety of memory types, including NVM SSDs and different DDR DRAM versions, providing significant flexibility in memory configurations. Industry proposals have explored integrating SSDs with CXL, with notable announcements including Intel's planned release of a CXL-attached Optane SSD and Samsung's introduction of the CXL hybrid SSD (CMM-H).

Although products based on these announcements have not yet been commercialized, it is expected that these SSDs will incorporate DRAM as a memory cache to mitigate the slower performance of the underlying storage media. Consequently, the performance of these SSD-based expanders is likely to depend heavily on the efficient management of their internal DRAM. In addition, write operations on such devices are anticipated to be slower than read operations and may occasionally encounter tail latency due to internal processes of the SSD controller. For instance, PRAM requires fine-grained wear-leveling, while low-latency flash memory necessitates garbage collection to reconcile the mismatch between write and erase unit sizes. Therefore, integrating SSD-based expanders with CXL will require careful management of write operations to ensure performance and reliability.

\begin{figure*}
  \centering
  \includegraphics[width=\linewidth]{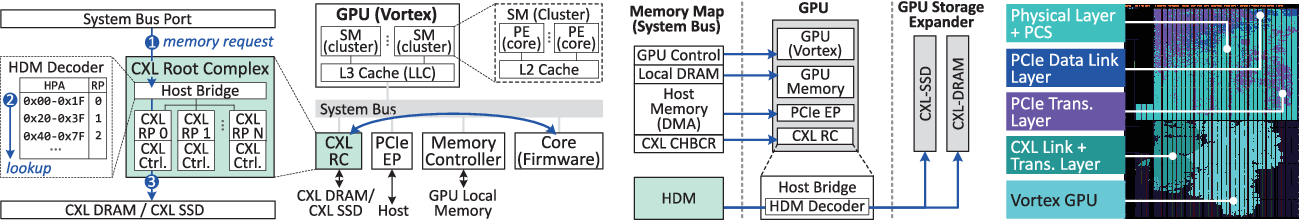}
  \begin{minipage}{\linewidth}
    \centering
    \vspace{0pt}
    \begin{subfigure}{1\linewidth}
      \renewcommand*{\arraystretch}{0.3}
      \begin{tabularx}{\linewidth}{
        p{\dimexpr.44\linewidth-1.33\tabcolsep}
        p{\dimexpr.31\linewidth-1.33\tabcolsep}
        p{\dimexpr.24\linewidth-1.33\tabcolsep}
        }
        \vspace{-0pt} \caption{GPU architecture.}\label{figs:configuration-hostbridge} &
        \vspace{-0pt} \caption{GPU memory space.}\label{figs:memoryspace} &
        \vspace{-0pt} \caption{Hardware prototypes.}\label{figs:floorplan-fpga}
      \end{tabularx}
    \end{subfigure}
    \vspace{5pt}
    \caption{Design of CXL-integrated GPU.}
  \end{minipage}
  \vspace{0pt}
\end{figure*}

\section{DESIGN OF CXL-INTEGRATED GPU}
\noindent \textbf{CXL hardware layer stack.}
In response to the lack of a publicly accessible, comprehensive hardware stack for CXL, we have developed a series of essential hardware layers that support the three key CXL subprotocols, consolidating these into a unified controller. Figure \ref{figs:floorplan-silicon} illustrates the silicon layout of our CXL controller, fabricated using advanced silicon technology. This controller is designed for compatibility with CXL 3.1 while also offering backward compatibility with earlier versions of CXL (2.0/1.1). The Flex Bus physical layer is integrated with our PCIe physical coding sublayer (PCS), allowing it to support both PCIe and CXL layer stacks seamlessly over elastic buffers \cite{cxl2023cxl}.

To address the dual requirements of PCIe and CXL, particularly in the context of power management and administrative operations, an arbitrator state machine has been incorporated into the controller. This state machine enables efficient resource allocation between PCIe and CXL tasks, ensuring both performance and system stability. The controller has undergone extensive testing, achieving a round-trip latency in the range of tens of nanoseconds, including the overhead associated with protocol conversion between standard memory operations and CXL flit-based transmissions.

This controller has been successfully integrated into hardware RTL implementations of both a memory expander and GPU/CPU prototypes, demonstrating its functionality and compatibility across diverse computing systems. For a clearer understanding of its performance, we compare the round-trip latency of our CXL controller against SMT \cite{kim2023smt} and TPP \cite{maruf2023tpp}. As shown in Figure \ref{figs:ctrl-perf}, our controller outperforms these alternatives, achieving a latency over three times faster. While publicly available details about the silicon design that SMT and TPP emulate (or partially leverage) are limited, we hypothesize that these controllers rely on PCIe architecture. In contrast, our controller is fully optimized for CXL, with tailored enhancements spanning the physical layer, link layer, and transaction layer.

\noindent \textbf{GPU architecture design and integration.}
In developing EP devices for GPU-based storage expansion, we have integrated the CXL controller with memory and SSD controllers, as depicted in the round-trip datapath in Figure \ref{figs:latency-analysis}. This integration allows the CXL controller to extend its backend storage to the host system as \emph{host-managed device memory} (HDM). The host is informed of this functionality through a feature analogous to PCIe capabilities, specifically adapted for HDM usage.

Integrating with GPU architectures, however, poses distinct challenges due to the need for the EP device to interface with the GPU's cache system. A direct implementation, similar to conventional EPs, is not viable in this scenario. To address these complexities, we have designed a specialized CXL root complex with a host bridge that includes multiple root ports. This configuration, shown in Figure \ref{figs:configuration-hostbridge}, connects the host bridge to the system bus port on one side and several CXL root ports on the other. A critical component of this setup is the HDM decoder, which manages the allocation of system memory address ranges, referred to as \emph{host physical addresses} (HPA), for each root port. These root ports are versatile, supporting DRAM or SSD-based EPs via PCIe connections. When a GPU compute unit issues a memory request (\blue{\circleNum{1}}), the corresponding root port translates the request into a CXL flit format (\blue{\circleNum{2}}) and forwards it to the CXL controller (\blue{\circleNum{3}}). This design ensures seamless communication between the GPU and the expanded storage, facilitating efficient data processing and access.

\begin{figure*}
  \begin{minipage}[t]{1\linewidth}
    \vspace{0pt}
    \begin{minipage}[t]{.24\linewidth}
      \vspace{0pt}
      \includegraphics[width=\linewidth]{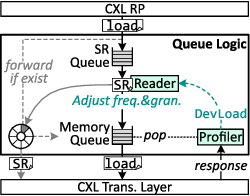}
    \end{minipage}%
    \begin{minipage}[t]{.5\linewidth}
      \vspace{0pt}
      \includegraphics[width=\linewidth]{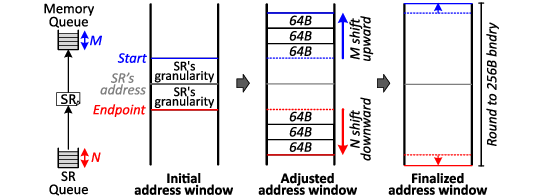}
    \end{minipage}%
    \begin{minipage}[t]{.24\linewidth}
      \vspace{0pt}
      \includegraphics[width=\linewidth]{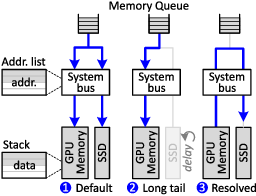}
    \end{minipage}
  \end{minipage}
  \begin{minipage}[t]{1\linewidth}
    \vspace{0pt}
    \begin{minipage}[t]{.26\linewidth}
      \vspace{20pt}
      \caption{Speculative read.}\label{figs:design-speculative-read}
    \end{minipage}%
    \begin{minipage}[t]{.45\linewidth}
      \vspace{20pt}
      \caption{Address window control.}\label{fig:address-window-control}
    \end{minipage}%
    \begin{minipage}[t]{.27\linewidth}
      \vspace{20pt}
      \caption{Deterministic store.}\label{figs:design-deterministic-store}
    \end{minipage}
  \end{minipage}
\end{figure*}

\noindent \textbf{System configuration and memory space.}
Figure \ref{figs:configuration-hostbridge} shows the placement of the CXL root complex within our GPU architecture, implemented using the Vortex GPU framework and its corresponding hardware RTL \cite{tine2021vortex}. In the Vortex architecture, the GPU\textquotesingle{}s computational units, known as streaming multiprocessors (SMs) or clusters, connect to the system bus through the last-level cache (LLC). The system bus also interfaces with the controller for GPU local memory and a PCIe EP, which supports host communication and kernel execution. In our design, the CXL root complex is integrated into the system bus alongside a simplified core responsible for initializing the connected EPs, the host bridge\textquotesingle{}s HDM decoder, and the HPAs associated with each root port. During this initialization phase, firmware identifies CXL EPs by examining their configuration space and PCIe BARs. It aggregates each EP\textquotesingle{}s memory address space by analyzing the HDM capability registers. The firmware then records this information in the HDM decoder of the host bridge, specifying the base address and size of the HDM for each root port (cf. Figure \ref{figs:configuration-hostbridge}).

Once initialization is complete, the GPU\textquotesingle{}s system bus memory map is structured as shown in Figure \ref{figs:memoryspace}. This map defines the allocation of address space across devices, segmented by function. For instance, when an SM generates a request targeting host memory, it interacts with the PCIe EP, which routes the request to the host. Similarly, GPU storage expanders managed by the CXL root complex are integrated into the system\textquotesingle{}s memory architecture. Upon receiving a memory request for this system memory segment, the CXL root complex consults the HDM decoder, converts the request into a CXL flit, and forwards it to the corresponding root port and controller.

The systems employing our CXL controllers, including the Vortex GPU and EP devices, are realized on a 7 nm FPGA-based custom add-in card (AIC), as shown in Figure \ref{figs:overview-and-framework2}. The associated RTL designs are detailed in Figure \ref{figs:floorplan-fpga}.

\section{OPTIMIZATION OF CXL CONTROLLER FOR GPUs}
Our CXL controller attains a round-trip latency in the two-digit range, though this latency varies depending on the type of backend media, such as DRAM or SSD. To address the latency challenges, particularly those associated with backend media access, we propose two optional strategies: (i) speculative read (SR) and (ii) deterministic store (DS). The SR strategy aims to minimize read latency, while the DS approach is designed to mitigate slower or tail latencies during write operations.

\subsection{Accelerating Reads with Speculation}
To enhance read performance, we employ the speculative read feature (\texttt{MemSpecRd}) introduced in CXL 2.0. This feature operates similarly to standard memory read or write requests but allows prefetching data from an address likely to be accessed in the near future. As shown in Figure \ref{figs:design-speculative-read}, our implementation incorporates queue logic beneath the root port, consisting of two separate queues: the SR queue and the memory queue, each with a capacity of 32 entries. When a load request is received, it is added to the SR queue, where a reader module processes the request and generates a speculative read (\texttt{MemSpecRd}) operation.

\noindent \textbf{Implementation.}
To adapt \texttt{MemSpecRd} for our implementation, we modify its address format. While originally designed for 64B granularity, we repurpose the two least significant bits to indicate the length of the request, with the remaining bits specifying a 256B memory offset. This adjustment enables the aggregation of details from multiple memory requests (ranging from 1 to 4) into a single \texttt{MemSpecRd} operation, improving efficiency and reducing latency.

If there is available space in the memory queue, an SR request generated by the SR reader module is transferred there. If no space is available, the request remains pending in the SR queue. Once in the memory queue, requests are forwarded to the CXL controller's transaction layer, ensuring they are processed by the target EP. In addition, the SR reader module records the address and length of each issued request in a ring buffer. If a new request matches the address of a previously issued SR request, it is directly forwarded as a standard memory request. This proactive transmission of load addresses allows the EP-side CXL controller to prefetch the corresponding data before receiving the actual read requests.

When the target EP responds, the queue logic's profiler removes the completed request from the memory queue. It is important to note that SR requests contribute to PCIe traffic. To manage this, the profiler assesses the EP's status by analyzing quality-of-service (QoS) telemetry data from the response, specifically the \texttt{DevLoad} field, as defined in the CXL flit. The \texttt{DevLoad} metric, which reflects the workload of the EP, is shared with the SR reader to dynamically adjust the frequency of SR requests. If the \texttt{DevLoad} value indicates a busier device, the SR request volume is reduced to avoid overloading the system. To further refine control, the queue logic determines the load of memory requests for each \texttt{MemSpecRd} and adjusts the queue length by referencing the \texttt{DevLoad} field returned from previous requests. This mechanism ensures effective management of SR requests and optimizes system performance. The details of this control and management strategy are described below.

\noindent \textbf{Load control for speculative reads.} Since the \texttt{DevLoad} field The \texttt{DevLoad} field, defined as two bits in the CXL standard, classifies the device\textquotesingle{}s workload into four states: light load (\texttt{ll}), optimal load (\texttt{ol}), moderate overload (\texttt{mo}), and severe overload (\texttt{so}). Each state dictates how the queue logic adjusts speculative read (SR) requests to match the available bandwidth and prevent unnecessary strain on the backend media.

When the device is in a \texttt{ll} state, the backend media has sufficient bandwidth to handle additional requests. In this case, the queue logic increases the granularity of SR requests by utilizing the flexibility of the \texttt{MemSpecRd} format. Specifically, it adjusts the request size from 256B to 1024B, enabling more efficient data prefetching. In the \texttt{ol} state, the device operates at full bandwidth capacity without being overwhelmed, and the queue logic maintains the current granularity of SR requests to preserve performance.

In contrast, a \texttt{mo} state indicates that the backend media is experiencing increased traffic with many outstanding requests. To alleviate this, the queue logic reduces the granularity of SR requests, ensuring that only essential data is fetched, thereby preventing further congestion. In a \texttt{so} state, the memory queue is completely saturated. The queue logic temporarily halts SR requests until the \texttt{DevLoad} field indicates a return to the light load state, allowing the system to recover. This adaptive load control mechanism dynamically adjusts SR request granularity and frequency to balance efficiency and system stability, ensuring optimal performance across varying workload conditions.

\noindent \textbf{Address window control.} SR can improve the read performance of SSD-based EP devices to match that of DRAM-based EPs by preloading large-granular data into the internal DRAM. However, because actual CXL requests operate at a 64B granularity, reading data in the wrong direction can result in internal DRAM pollution. This pollution occurs when unnecessary data occupies internal DRAM capacity and consumes bandwidth between the DRAM and backend media. For instance, if a user program accesses an array in reverse order, and the queue logic reads data forward from the SR request\textquotesingle{}s address according to the granularity, it may lead to substantial pollution of the internal DRAM.

To prevent such inefficiencies, the queue logic determines an optimal address window for each incoming SR request by analyzing both the memory queue and SR queue, as shown in Figure \ref{fig:address-window-control}. An address window is defined as the range of addresses starting from a calculated offset and extending to an endpoint determined by the SR request\textquotesingle{}s granularity. The queue logic begins by setting an initial address window. The starting offset is calculated as the incoming SR request\textquotesingle{}s address minus its granularity, while the endpoint is the SR request\textquotesingle{}s address plus its granularity. Since requests in the memory queue represent prior requests and those in the SR queue represent anticipated future requests, the queue logic adjusts the initial window accordingly. For each request in the memory queue, the queue logic shifts the start of the address window upwards by the CXL memory request granularity (64B). In contrast, for each request in the SR queue, the logic shifts the endpoint of the address window downwards by the same granularity.

Once all requests have been evaluated, the adjusted address window is finalized by rounding the shifted range to the nearest 256B boundary, which corresponds to the memory offset unit for an SR request. This ensures that SR requests target the most relevant data range, minimizing unnecessary data movement and optimizing internal DRAM utilization.

\subsection{Deterministic Store for Precise Writes}
The proposed DS technique offers a solution to conceal such characteristics by leveraging reserved space in GPU memory.

\noindent \textbf{Design and implementation.}
As shown in Figure \ref{figs:design-deterministic-store}, when a write operation to an SSD is initiated in the memory queue, the request is concurrently sent to both the GPU memory and the SSD (as orchestrated by the system memory map), allowing for the immediate release of the request (\blue{\circleNum{1}}). Should there be a delay observed from the SSD prior to the arrival of the subsequent write request, the queue logic temporarily stores the data in the GPU memory's reserved address (\blue{\circleNum{2}}).
This temporary data storage is organized using a stack structure, designed to collapse in the event of detected tail latency or slow writes. On the other hand, if no such issues are observed, the data within the GPU memory at the designated address is updated, reflecting the dual write operation (GPU memory/SSD). To maintain a record of each stack entry's precise location, an address list is stored within the system bus's internal SRAM. The stack is methodically emptied by flushing the data in the background (\blue{\circleNum{3}}). Through the implementation of DS, from the perspective of the SMs and their LLC controller, the CXL root complex exhibits deterministic store characteristics, effectively shielding them from variations in write latency.

\noindent \textbf{Fine control for internal tasks.}
While the proposed hybrid memory management reduces the variability in write latency, internal SSD tasks such as garbage collection require finer control to mitigate their long tail latency. These tasks can temporarily make the SSD unavailable and cause congestion at the ingress port, degrading the performance of even read operations. To address this, we utilize the \texttt{DevLoad} field, as specified in the CXL standard, in a manner tailored for writes as well. When an internal task is expected to reduce throughput temporarily, the backend media reports this condition through the \texttt{DevLoad} field before scheduling the task. By monitoring \texttt{DevLoad}, our controller dynamically throttles write requests to the affected port, ensuring that the temporary throughput reduction does not lead to severe performance degradation.

The controller for each root port continuously tracks the state of the storage media by monitoring changes in \texttt{DevLoad}. If an increase in \texttt{DevLoad} is detected, indicating congestion or an internal task, the controller temporarily suspends issuing new write requests to the affected port. Instead, incoming requests are forwarded to GPU memory to maintain system throughput. During this period, the controller keeps the states in the system bus's internal SRAM, which is implemented as a red-black tree for efficient management.

The controller periodically checks the \texttt{DevLoad} value. Once the reported load decreases, signaling that the internal task has completed or congestion has eased, the controller resumes the suspended writes by issuing them to the target port. For read operations, the controller first checks the system bus buffer. If the requested data is present in the buffer, the read is served directly from GPU memory, bypassing the backend media. This approach ensures minimal disruption to read operations while maintaining effective control over write requests, thereby optimizing overall system performance.


\section{EVALUATION}

\begin{figure*}
  \centering
  \includegraphics[width=\linewidth]{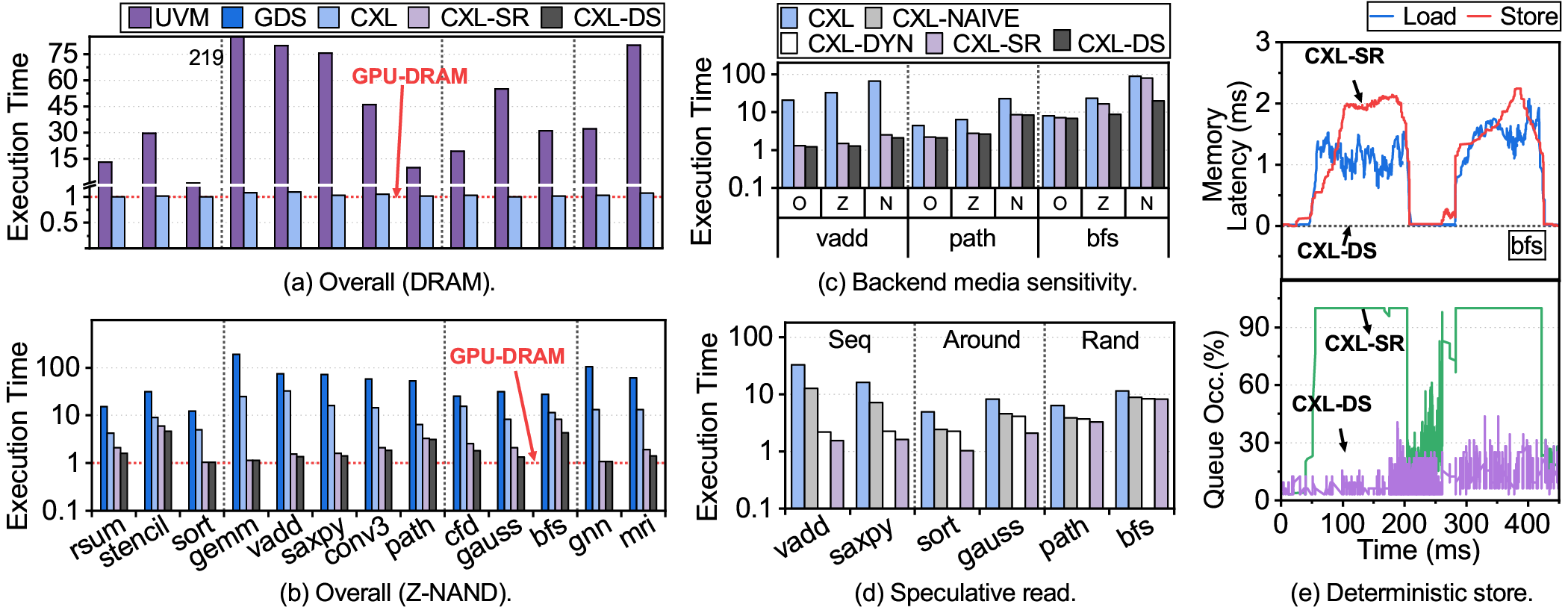}
  \vspace{-5pt}
  \caption{Performance analysis.}\label{fig:eval}
  \vspace{0pt}
\end{figure*}

\begin{table}
  \begin{subfigure}[t]{.4\linewidth}
    \vspace{0pt}
    \setlength\tabcolsep{2.4pt}
    \renewcommand{\arraystretch}{1.1}
    \resizebox{.98\linewidth}{!}{
    \begin{tabular}[c]{@{}lr@{}}
      \toprule
      \multicolumn{2}{c}{\textbf{Vortex Configuration}} \\
      \multicolumn{2}{c}{Cores/Threads 8/8} \\
      \midrule
      \multicolumn{2}{c}{\textbf{PCIe Configuration}} \\
      \multicolumn{2}{c}{PCIe 5.0 (32GT/s) x8} \\
      \midrule
      \multicolumn{2}{c}{\textbf{CXL Configuration}} \\
      \multicolumn{2}{c}{PCIe 5.0 (32GT/s) x8} \\
      \multicolumn{2}{c}{Sync-header bypass} \\
      \midrule
      \textbf{DRAM} & DDR5-5600\cite{li2020dramsim3} \\
      \midrule
      \textbf{Optane} & Intel P5800X \\
      \midrule
      \shortstack{\textbf{Z-NAND} \\ \,} & \shortstack{Samsung \\ 983 ZET} \\
      \midrule
      \shortstack{\textbf{NAND} \\ \,} & \shortstack{Samsung \\ 980 Pro} \\
      \bottomrule
    \end{tabular}
    }
    \caption{Configs.}\label{tbl:configs}
  \end{subfigure}%
  \hspace{.02\linewidth}
  \begin{subfigure}[t]{.577\linewidth}
    \vspace{0pt}
    \setlength\tabcolsep{2.4pt}
    \renewcommand{\arraystretch}{1.1}
    \resizebox{.98\linewidth}{!}{
    \begin{tabular}[c]{@{}cccc@{}}
      \toprule
      \shortstack{\textbf{Workload} \\ \textbf{Type}}
        &  \shortstack{\textbf{Acronym} \\ \,}
        & \shortstack{\textbf{Compute} \\ \textbf{Ratio}}
        & \shortstack{\textbf{Load} \\ \textbf{Ratio}} \\
      \midrule
      \multirow{3}{*}{\shortstack{Compute \\ Intensive}}
        & \texttt{rsum}     & 31.4\% & 53.3\% \\
        & \texttt{stencil}  & 37.5\% & 72.5\% \\
        & \texttt{sort}     & 38.1\% & 98.7\% \\
      \midrule
      \multirow{5}{*}{\shortstack{Load \\ Intensive}}
        & \texttt{gemm}    & 11.6\% & 99.9\% \\
        & \texttt{vadd}     & 15.6\% & 69.1\% \\
        & \texttt{saxpy}    & 16.2\% & 69.2\% \\
        & \texttt{conv3}    & 21.8\% & 78.6\% \\
        & \texttt{path}     & 27.0\% & 92.7\% \\
      \midrule
      \multirow{3}{*}{\shortstack{Store \\ Intensive}}
        & \texttt{cfd}      & 20.9\% & 42.6\% \\
        & \texttt{gauss}    & 23.5\% & 48.5\% \\
        & \texttt{bfs}      & 29.3\% & 43.2\% \\
      \midrule
      \multirow{2}{*}{\shortstack{Real-world \\ Workload}}
        & \texttt{gnn}      & 27.4\% & 73.8\% \\
        & \texttt{mri}    & 29.2\% & 53.3\% \\
      \bottomrule
    \end{tabular}
    }
    \caption{Workloads.}\label{tbl:workloads}
  \end{subfigure}
  \vspace{15pt}
  \caption{Evaluation setup.}
  \vspace{0pt}
\end{table}

\noindent \textbf{Methodology.}
Our hardware prototype provides detailed latency characteristics that are not accessible to end users but lacks the flexibility required to explore the design space for GPU storage expansion. To address this limitation, we developed a simulator that accurately models the behavior of our prototype and used it for evaluation purposes. The simulator reproduces the RTL logic\textquotesingle{}s behavior by leveraging data extracted from real-world workload execution and RTL behavior simulations. For example, we utilized execution times and GPU pipeline statistics obtained from Vortex\textquotesingle{}s performance counters and waveform dumps. Memory latencies for the Vortex GPU and CXL endpoints were calculated using DRAMSim3 \cite{li2020dramsim3}.

In addition to DRAM, we evaluated various storage media, including PRAM (Optane), ultra-low-latency flash (Z-NAND), and conventional flash (NAND). The key characteristics of these storage types, critical for accurate simulation, are summarized in Table \ref{tbl:configs}. For PCIe and CXL bus latencies, we used real values measured from our ASIC. Further details of our simulation setup are also provided in Table \ref{tbl:configs}. In addition, UVM and GPUDirect contribute to host runtime overhead, which we account for as approximately 500 $\mu s$, based on the findings reported in \cite{allen2021demystifying}.

\noindent\textbf{Configurations.} For the evaluation, we consider five different GPU configurations: i) \texttt{UVM}, ii) \texttt{GDS}, iii) \texttt{CXL}, iv) \texttt{CXL-SR}, and v) \texttt{CXL-DS}. \texttt{UVM} employs NVIDIA's unified virtual memory \cite{nvidia2022cuda}, which extends GPU memory by utilizing host DRAM. \texttt{GDS} uses GPUDirect storage \cite{nvidia2022gpudirectstorage}, enabling direct connections between the GPU and underlying storage. In both cases, data is transferred to the GPU per page through the host runtime.

The \texttt{CXL} configuration represents an expander implemented with the optimized controller proposed in this paper. The \texttt{CXL-SR} and \texttt{CXL-DS} configurations build upon \texttt{CXL} by incorporating speculative read (SR) and deterministic store (DS) techniques, respectively, to further enhance performance.

We also evaluate an ideal configuration, \texttt{GPU-DRAM}, which assumes sufficient on-device GPU memory and eliminates the need for any host-side memory expansion. For SR and DS evaluations, we simulate different types of backend storage media, including Optane, Z-NAND, and NAND, with their model configurations detailed in Table \ref{tbl:configs}.

\noindent \textbf{Workloads.} We evaluate eleven workloads selected from the Rodinia benchmark suite, which includes a diverse set of parallel programs designed for GPUs \cite{che2009rodinia}.
We also evaluate two real-world workloads, \texttt{gnn} and \texttt{mri}. The \texttt{gnn} workload is composed of \texttt{bfs}, \texttt{vadd}, and \texttt{gemm}, while the \texttt{mri} workload is composed of \texttt{sort} and \texttt{conv3}.
To ensure a fair comparison across all GPU configurations, the input data sizes are adjusted to fit within the 10$\times$ bigger capacity of the GPU's local memory. For clarity, the workloads are categorized into compute-intensive and memory-intensive applications, with the latter further divided into load-intensive and store-intensive subsets. All workloads are arranged in ascending order based on their memory access ratios to facilitate analysis and comparison.

\subsection{Performance Analysis}
Figures \ref{fig:eval}\textcolor{blue}{a} and \ref{fig:eval}\textcolor{blue}{b} present the performance of various GPU configurations with CXL memory expanders using DRAM and Z-NAND as backend media, respectively. All execution times are normalized to the performance of the \texttt{GPU-DRAM} configuration.

\noindent \textbf{DRAM-based expanders.}
Figure \ref{fig:eval}\textcolor{blue}{a} compares \texttt{UVM}, \texttt{CXL}, and \texttt{GPU-DRAM}, as SR and DS mechanisms are only relevant for expanders with non-DRAM backend media. As shown, \texttt{UVM} performs 52.7$\times$ worse than \texttt{GPU-DRAM}, with particularly pronounced latency increases for load-intensive workloads such as \textit{gemm}, \textit{vadd}, and \textit{saxpy}. These workloads primarily involve sequential operations like vector or scalar multiplication, where data is read once and seldom accessed again. This access pattern generates numerous page faults, causing significant delays due to host runtime interventions.

In contrast, \texttt{CXL} reduces latency substantially compared to \texttt{UVM}, achieving a performance improvement of 44.2$\times$ by accessing the extended DRAM directly without host runtime involvement. The performance of \texttt{CXL} approaches that of \texttt{GPU-DRAM}, being slower by only 2.3\%, 19.7\%, and 6.8\% for compute-intensive, load-intensive, and store-intensive applications, respectively. Compute-intensive workloads exhibit significantly fewer memory accesses compared to load-intensive ones (87.5\% less), and most of these accesses are cache hits, making \texttt{CXL} nearly indistinguishable from \texttt{GPU-DRAM}. For store-intensive applications, the impact of \texttt{CXL} is less pronounced compared to load-intensive workloads. This is because store-intensive applications tend to buffer data extensively, mitigating the effects of memory access latencies. As a result, the performance gap between \texttt{CXL} and \texttt{GPU-DRAM} is relatively small in these scenarios.
Note that the real-world applications show patterns similar to the operations that compose each application.

\noindent \textbf{SSD-based expanders.} Due to the significant performance variation across the GPUs evaluated, Figure \ref{fig:eval}\textcolor{blue}{b} presents the results using a logarithmic scale. This analysis focuses on the benefits of the SR (Speculative Read) and DS (Deterministic Store) techniques.

As shown, \texttt{CXL-SR} achieves an average performance improvement of 7.4$\times$ compared to \texttt{CXL}. This improvement is attributed to the CXL controller and queue logic, which use SR requests to preload data into internal DRAM, allowing sequential reads to be served immediately. However, the effectiveness of SR depends on applications exhibiting specific data access localities (256B to 1024B) and avoiding excessive jumps across different SR requests. In addition, the internal DRAM is a constrained resource, which requires access patterns to remain relatively consistent.

For instance, 1D vector computation algorithms such as \textit{vadd} and \textit{saxpy} exhibit the highest performance gains (15.6$\times$) because they issue SR requests for upcoming computations while processing the current data. Similarly, 2D array computation workloads (e.g., \textit{gemm}, \textit{conv3}) also benefit from SR's locality-based optimizations. In contrast, graph-based applications such as \textit{path} and \textit{bfs}, which feature irregular access patterns, show limited benefits (67.6\%, on average). Although finer-granular SR requests are possible, internal DRAM can become crowded, leading to speculative data eviction and reducing the effectiveness of SR.

On the other hand, \texttt{CXL-DS} delivers average performance improvements of 20.9\%, 8.7\%, and 62.8\% over \texttt{CXL-SR} for various workload categories. These benefits are particularly pronounced in store-intensive applications, as DS mitigates tail latency caused by internal SSD operations such as garbage collection. The deterministic store mechanism effectively hides such latency behind normal operations, allowing GPU threads to continue execution without prolonged suspensions during store operations. Detailed analysis of the long-tail latency mitigation will be discussed later.

\subsection{Backend Media Latency Mitigation}

Figure \ref{fig:eval}\textcolor{blue}{c} illustrates the performance benefits of the SR and DS mechanisms across various CXL endpoints' backend memory media, specifically Optane (\texttt{O}), Z-NAND (\texttt{Z}), and standard NAND (\texttt{N}).
We analyzed three workloads: one with a sequential, read-intensive access pattern (\texttt{vadd}) and two with random access patterns, each  load-intensive (\texttt{path}) and store-intensive (\texttt{bfs}), respectively.

SR hides the latency of the backend media through preloading, providing an average performance improvement of 7.1$\times$, 8.8$\times$, and 10.1$\times$ for Optane, Z-NAND, and standard NAND, respectively.
However, the extent of the performance gain varies depending on the proportion of preloaded data that is actually accessed.
For instance, applications like \texttt{vadd} achieves only 2.4$\times$ slowdown in \texttt{N} configuration compared to \texttt{GPU-DRAM}.
\texttt{path} and \texttt{bfs} also seek 2.4$\times$ and 1.2$\times$ improvement, respectively, in average.

DS offers additional performance improvements for store-intensive applications by hiding the tail latency caused by media mangement such as garbage collection (GC).
Consequently, it provides up to an 4$\times$ performance gain for \texttt{bfs}.
We further discuss how the proposed SR and DS mechanisms deliver such performance benefits.

\noindent\textbf{Speculative read.}

Figure \ref{fig:eval}\textcolor{blue}{d} illustrates the performance benefits of each optimization applied to the proposed SR mechanism. The evaluation was conducted using a Z-NAND-based system while incrementally adding optimization techniques to the SR mechanism.
The configuration \texttt{CXL-NAIVE} builds on \texttt{CXL} by implementing a naive SR unit that blindly issues 64B \texttt{MemSpecRd} commands about all memory requests. By contrast, \texttt{CXL-DYN} extends \texttt{CXL-NAIVE} by utilizing the lower two bits of the memory address to issue larger SR requests. Using the \texttt{DevLoad} field of the endpoint device, \texttt{CXL-DYN} determines the size of SR requests, while retaining the starting address of the original memory request.
The configuration \texttt{CXL-SR} further enhances \texttt{CXL-DYN} by analyzing the requests in the SR and memory queues in order to dynamically determine the starting address of SR requests at runtime.

To evaluate the effectiveness of these configurations, the workloads were categorized into three primary memory access patterns: \texttt{Seq}, \texttt{Around}, and \texttt{Rand}.
The \texttt{Seq} and \texttt{Rand} patterns represent sequential and random memory accesse, and were evaluated using 1D vector algorithms and graph-based algorithms, respectively.
The \texttt{Around} pattern describes access patterns of which spatial locality exists but the direction of access may not be monotonically increasing.
This pattern was examined using binary tree algorithms such as \texttt{sort}, which access either the left or right child node based on the comparison results, and Gaussian elimination (\texttt{gauss}), where runtime decisions determine whether to access the current or previous row based on the first element of the row.

\texttt{CXL-NAIVE} increases performance 1.9$\times$ compared to \texttt{CXL}.
This benefit occurs because data can be read from NAND in advance by sending \texttt{MemSpecRd} requests for requests that are waiting in the GPU's memory queue. This approach prevents cold misses in the SSD's internal DRAM for all workloads. Specifically, under \texttt{CXL}, the \texttt{Seq}, \texttt{Around}, and \texttt{Rand} workloads reach DRAM hit rates of 47.4\%, 31.2\%, and 10\%, respectively, while under \texttt{CXL-NAIVE}, they reach 88.4\%, 56\%, and 32.1\%, respectively.

However, \texttt{CXL-NAIVE} remains 6.6$\times$ slower than \texttt{GPU-DRAM} due to the limited coverage of the SR requests.
\texttt{CXL-DYN} achieves an additional performance gain of 4.5$\times$ for \texttt{Seq} workloads by providing the opportunity to preload data more data promptly.
This is because the aggressive preloading raises the SSD DRAM hit rate for \texttt{Seq} workloads to over 99\%.
\texttt{Around} and \texttt{Rand} workloads gain less due to the probabilistic utilization of the preloaded data, resulting in 10.6\% and 5.7\% improvement, respectively.
For \texttt{Rand} workloads, simply increasing the granularity of preloading can lead to SSD DRAM pollution and reduce performance. In practice, sending the largest possible SR requests at all times lowers the SSD DRAM hit rate for \texttt{Rand} from 32.1\% to 30.7\%. However, since \texttt{CXL-DYN} can dynamically adjust the granularity of \texttt{MemSpecRd} requests, it achieves a higher hit rate (34\%) for \texttt{Rand}.

For \texttt{Around} workloads, which generally follow sequential access patterns, the SSD DRAM hit rate is higher than \texttt{CXL-NAIVE}, but it still remains at 57.4\% because the next access may occur either before or after the current address. \texttt{CXL-SR} overcomes this limitation by dynamically adjusting the address window of the SR requests, yielding an additional 2.1$\times$ performance improvement for \texttt{Around}. This is possible because \texttt{CXL-SR} can analyze the incoming addresses and decide whether to send \texttt{MemSpecRd} requests for addresses before or after the current one, raising the SSD DRAM hit rate for \texttt{Around} to 75.8\%.

\balance

\noindent\textbf{Deterministic store.}

Figure \ref{fig:eval}\textcolor{blue}{e} illustrates the time-series of load/store request latencies and the utilization of the endpoint's ingress queue, for a write-intensive workload (\texttt{bfs}).
For the evaluation, we employed a Z-NAND-based system and compared the performance between a system implementing the proposed DS mechanism (\texttt{CXL-DS}) and one without it (\texttt{CXL-SR}), after a garbage collection (GC) is initiated.
For \texttt{CXL-SR}, the increased write latency immediately fills the ingress queue with the delayed write requests, resulting in a significant rise in both load and store latencies within a short period.
The performance degradation is critical, considering that it recurs shortly after GC completion.
This is because, even when the free blocks are temporarily secured through GC, the accumulated write requests in the ingress queue flood back to the memory media, triggering the next GC.
In contrast, \texttt{CXL-DS} forwards the store requests only to the GPU's local memory during media management periods, preventing them from being sent to the endpoint.
This strategy not only hides the tail latency of write requests but also prevents the tail latency of read requests by avoiding congestion in the ingress queue.
Moreover, by processing load and store requests in the GPU's local memory, it further reduces read latency and prevents the recurrence of frequent GCs.


\section{CONCLUSION}
We introduce a method to increase GPU storage using CXL technology, integrating DRAM and/or SSDs through CXL multiple root ports.
Our custom-designed CXL controller, implemented directly in the hardware, achieves fast response times not seen before in this field.
We further refine this system with features designed to improve handling of both read and write operations to storage media.
Our tests show that this approach improves upon existing solutions, offering a significant step forward in GPU storage capacity and efficiency.


\section{ACKNOWLEDGEMENT}
\label{sec:acknowledgement}
This work is protected by one or more patents.
The authors would like to thank the anonymous
reviewers for their comments, and Myoungsoo Jung is the
corresponding author (mj@panmnesia.com).

\def\refname{REFERENCES}

\bibliographystyle{IEEEtran}

\bibliography{IEEEabrv,ref}

\end{document}